\documentclass[twocolumn,aps,prl]{revtex4}

\usepackage{amsmath,amsfonts,amssymb}
\usepackage{wrapfig}
\usepackage{graphicx}
\usepackage{bbm}
\usepackage{color}

\usepackage[english]{babel}
\makeatletter

\def\gsim{\compoundrel>\over\sim}

\def\compoundrel#1\over#2{\mathpalette\compoundreL{{#1}\over{#2}}}
\def\compoundreL#1#2{\compoundREL#1#2}
\def\compoundREL#1#2\over#3{\mathrel
      {\vcenter{\hbox{$\m@th\buildrel{#1#2}\over{#1#3}$}}}}
\makeatother

\usepackage{babel}
\makeatother

\def\be{\begin{eqnarray}}
\def\ee{\end{eqnarray}}
\def\bee{\begin{eqnarray*}}
\def\eee{\end{eqnarray*}}

\begin{document}

\title{Dynamic entanglement in oscillating molecules and potential biological implications}

\author{Jianming Cai$^{1,2}$, Sandu Popescu$^{3,4}$ and Hans J. Briegel$^{1,2}$}

\affiliation{$^1$Institut f{\"u}r Theoretische Physik,
Universit{\"a}t Innsbruck, Technikerstra{\ss }e 25, A-6020 Innsbruck\\
$^2$ Institut f{\"u}r Quantenoptik und Quanteninformation der
\"Osterreichischen Akademie der Wissenschaften, Innsbruck, Austria\\
$^3$ H.H. Wills Physics Laboratory, University of Bristol, Tyndall
Avenue, Bristol BS8 1TL, U.K.\\
$^4$ Hewlett-Packard Laboratories, Stoke Gifford, Bristol BS12 6QZ,
U.K.}

\begin{abstract}
We demonstrate that entanglement can persistently recur in an
oscillating two-spin molecule that is coupled to a hot and noisy
environment, in which no static entanglement can survive. The system
represents a non-equilibrium quantum system which, driven through
the oscillatory motion, is prevented from reaching its (separable)
thermal equilibrium state. Environmental noise, together with the
driven motion, plays a constructive role by periodically resetting
the system, even though it will destroy entanglement as usual. As a
building block, the present simple mechanism supports the
perspective that entanglement can exist also in systems which are
exposed to a hot environment and to high levels of de-coherence,
which we expect e.g. for biological systems. Our results furthermore
suggest that entanglement plays a role in the heat exchange between
molecular machines and environment. Experimental simulation of our
model with trapped ions is within reach of the current
state-of-the-art quantum technologies.
\end{abstract}

\maketitle

The question, to what extent quantum mechanics
plays a role in biology, is still far from being well-understood
\cite{Abbott08,Briegel0806}. It seems that classical concepts alone
are insufficient for a proper understanding of certain biological
processes, and that coherent quantum effects need to be taken into
account. It has e.g.\ long been known that quantum tunneling plays an
important role in enzymatic reactions \cite{Ball04,Tunnelling}.
Experimental evidence for quantum coherence in the photosynthetic system
has recently been reported in \cite{Fleming07}. The interplay between the
coherent free Hamiltonian and the environment is believed to significantly
enhance quantum transport in the Fenna-Matthews-Olson (FMO) protein complex
\cite{Mohseni0805,Rebentrost0807,Plenio0807}.

Apart from these isolated instances where quantum coherence seems to
help, it is however not clear to what extent biological systems
exploit quantum mechanics e.g. to optimize their functionality
(beyond the trivial fact that the latter determines, of course, the
structure of bio-molecules). Most physicists and biologists are
generally skeptical about the question whether genuine quantum
features such as entanglement play a broader role in biology. The
obvious reason for that viewpoint is that entanglement is very
sensitive to noise and requires special conditions to be maintained,
in particular very good insulation. Biological systems are anything
but - they are wet and hot, and with extremely high levels of noise.

An often ignored fact is, however, that biological systems are also
open driven quantum systems, operating far away from thermal
equilibrium \cite{Briegel0806}. This opens many new possibilities which
have not yet been carefully considered. Different from e.g. solid
state physics, things in biology {\em move}. Protein functions, for
example, require conformational motion \cite{Frauenfelder99,Alberts08}.
During such motion, e.g. in the context of protein folding or in isomeric
transitions, we have to consider time-dependent quantum interactions
(capable of forming e.g. hydrogen or ionic bonds) which are
effectively switched on and off while the molecule changes its
shape.  Since these interactions are accompanied by a substantial
amount of noise (e.g. from fluctuating dipole fields from the
hydration shell and the bulk solvent), they are usually treated
classically. A proper understanding of protein dynamics may however
require one to explore the capability of these time dependent
interactions and whether they need to be treated quantum mechanically
\cite{Gilmore08}. It is, for example, not clear whether or not
entanglement is generated during these motional processes. A
positive answer to this question might reveal novel and subtle
aspect of protein dynamics and bio-motoric processes. It would also
provide a new twist to the study of nano-bio interfaces, which are
just in their infancy.

\begin{figure}[htb]
\begin{center}
\includegraphics[width=7cm,clip]{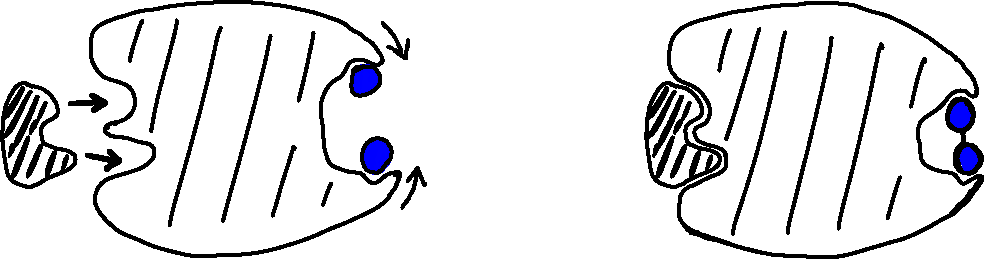}
\end{center}
\caption{Conformational changes of a bio-molecule \cite{Alberts08},
induced e.g. by the interaction with some other chemical, can lead to
a time-dependent interaction between different sites (blue)
of the molecule. See also \cite{Briegel0806}.}
\label{AllostericDevice}
\end{figure}

While it seems reasonable to describe the conformational motion of a
bio-molecule classically, it also carries quantum degrees of freedom,
such as nuclear spins and electronic states. These give rise
to effectively time-dependent quantum interactions, with their strengths
depending on the molecular shape, see Fig.~\ref{AllostericDevice}.
We analyze the role of environmental noise and decoherence in such
interactions. We find that entanglement can be generated even at room
temperature and despite the presence of decoherence, suggesting that
the underlying interactions should indeed be described fully quantum
mechanically to account for more subtle processes.

\begin{figure}[htb]
\begin{center}
\includegraphics[width=7cm,clip]{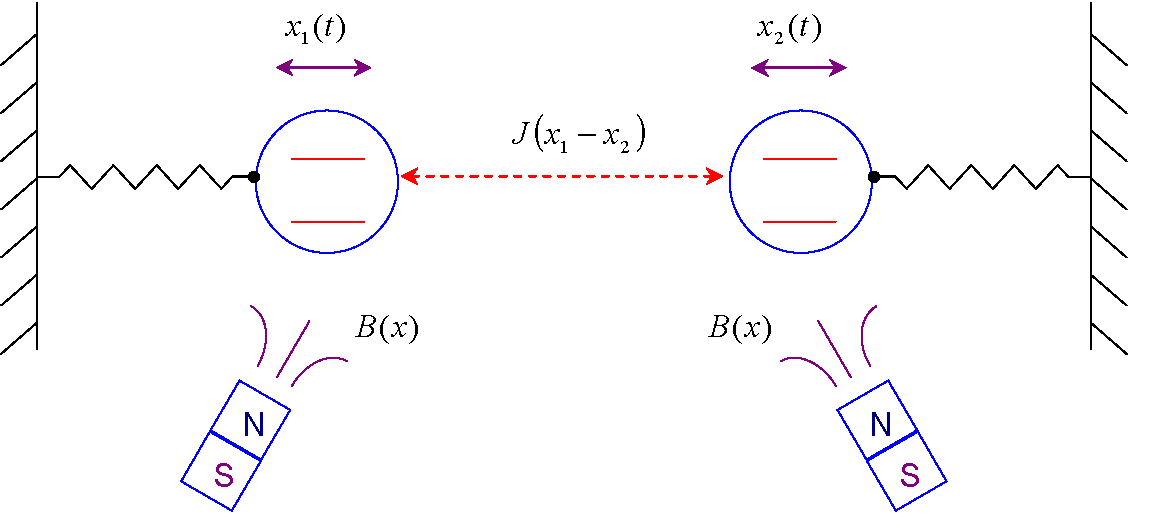}
\end{center}
\caption{Model of a two-spin molecule which undergoes conformational changes as
a function of time. Both the spin-spin interaction strength $J$ and local
fields $B$ are position dependent.}
\label{Conf}
\end{figure}

As a paradigmatic example, we study the time-dependent interaction of two spins
in a thermal and decoherent environment. We may imagine that the spins are
attached to some classical backbone structure whose shape changes in time,
as drawn schematically in Fig.~\ref{Conf}. For simplicity we call such an
arrangement a {\em two-spin molecule}.
We demonstrate that, if the distance between the spins is oscillating, cyclic
generation of fresh entanglement can persist, even if no static entanglement
can survive.  Environmental noise plays thereby both a destructive and constructive
role by effectively resetting the system \cite{Hartmann06}. The oscillating
molecule may be viewed as a molecular machine that exchanges heat with the reservoir.
Our results then suggest that entanglement is relevant to the absorption of
heat from the environment, which might possibly affect certain biological
processes.

In the semi-quantal picture that we
have introduced above, the conformational changes lead to {\em classical
motion of quantum degrees of freedom} as illustrated in Fig.~\ref{Conf}.
We assume that the two spins are coupled with Ising interaction and
that there also exist local electric and/or magnetic fields, both
of which are usually position dependent. Thus, the classical
molecular motion induces an effective time-dependent
Hamiltonian of the form
\begin{equation}
H_{M}(t)=J(t)\sigma^{(1)}_{x}\sigma^{(2)}_{x}+B(t)(\sigma^{(1)}_{z}+\sigma^{(2)}_{z})
\label{Hamiltonian}
\end{equation}
where $\sigma _{x}^{(\alpha )}$ and $\sigma _{z}^{(\alpha )}$ are
Pauli operators of the $\alpha $th spin, $J(t)$ is the interaction strength,
and $\omega_{0}(t)=2B(t)$ the local level splitting.
We emphasize that the subsequent results also hold for more general
Hamiltonians, but for simplicity we concentrate here on the Ising
interaction. The coupling of the spins to the environment will be
described by a master equation of the form
\begin{equation}
\frac{\partial}{\partial t}\rho(t)=-i[H_{M}(t),\rho ]+\mathcal{D}\rho(t) \equiv
{\mathcal L}(t) \rho(t)
\label{MasterEquation}
\end{equation}
where $\mathcal{D}\rho=\sum_{\mu}2L_{\mu}\rho
L_{\mu}^{\dagger}-L_{\mu}^{\dagger} L_{\mu}\rho-\rho L_{\mu}^{\dagger} L_{\mu}
$ describes the effect of the molecule-environment coupling, and
$L_{\mu}$ are Lindblad-type generators. 


The effect of the environment on the motion of bio-molecules is complex and far from being understood. To demonstrate the essential physics, we first consider a worst-case scenario, where the environment is described by bosonic heat-bath, with each spin being coupled to an independent thermal bath of harmonic oscillators \cite{BreuerBook,Gilmore07}.
In the static case, it is well-known that, above a certain temperature, no initial entanglement can persist in such an environment. We will however show that, even under such unfavorable conditions, entanglement can be generated if the particles start oscillating and the system moves out of equilibrium.

If the molecular oscillation is not too fast, in the sense that the adiabatic condition for closed systems is satisfied, the effect of the environment on the oscillating molecule can be described by a master equation of type (\ref{MasterEquation}), with implicitly time-dependent Lindblad generators (see also Supplementary Information).
As far as the {\em static entanglement} is concerned, at every
molecular configuration (with {\em fixed} $J$ and $B$), the molecule
will be driven towards its thermal equilibrium state
$\rho_{th}$ at temperature $T$. In the following, we adopt the concurrence ${\cal C}(\rho)$ \cite{Wootters98} as the measure of two-qubit entanglement of a state $\rho$. For a separable (non-entangled) state it vanishes, while for a maximal entangled state it reaches the value 1, i.e. $0\le {\cal C}(\rho) \le 1$. It can be shown that, if the temperature is above a critical value $T_{c}$, no entanglement can survive in any static configuration of the molecule, i.e. ${\cal C}(\rho_{th})=0$. In the following, we will only consider such situations where $T>T_{c}$.

The main question we are asking is this: {\em Can entanglement possibly build-up through the classical motion of the molecule?} The answer is affirmative and we demonstrate that entanglement can indeed persistently recur in an oscillating molecule, even if the environment is so hot that the static thermal state is separable for all possible molecular configurations, i.e. $T> \max\{T_{c}\}$.

\begin{figure}[htb]
\begin{center}
\begin{minipage}{8.8cm}
\hspace{-0.1cm}
\includegraphics[width=4.1cm]{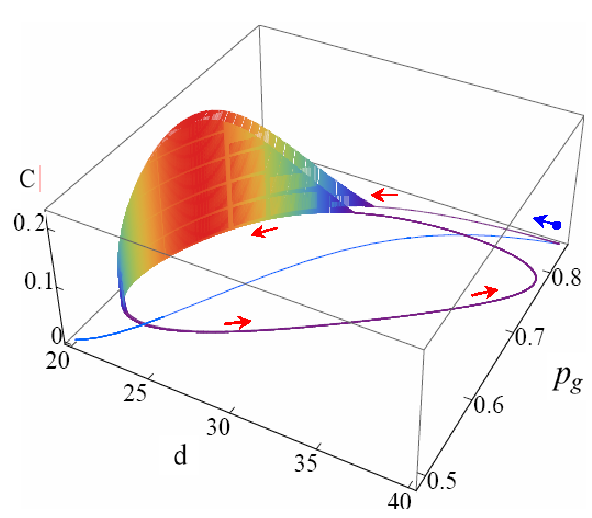}
\hspace{0cm}
\includegraphics[width=4.5cm]{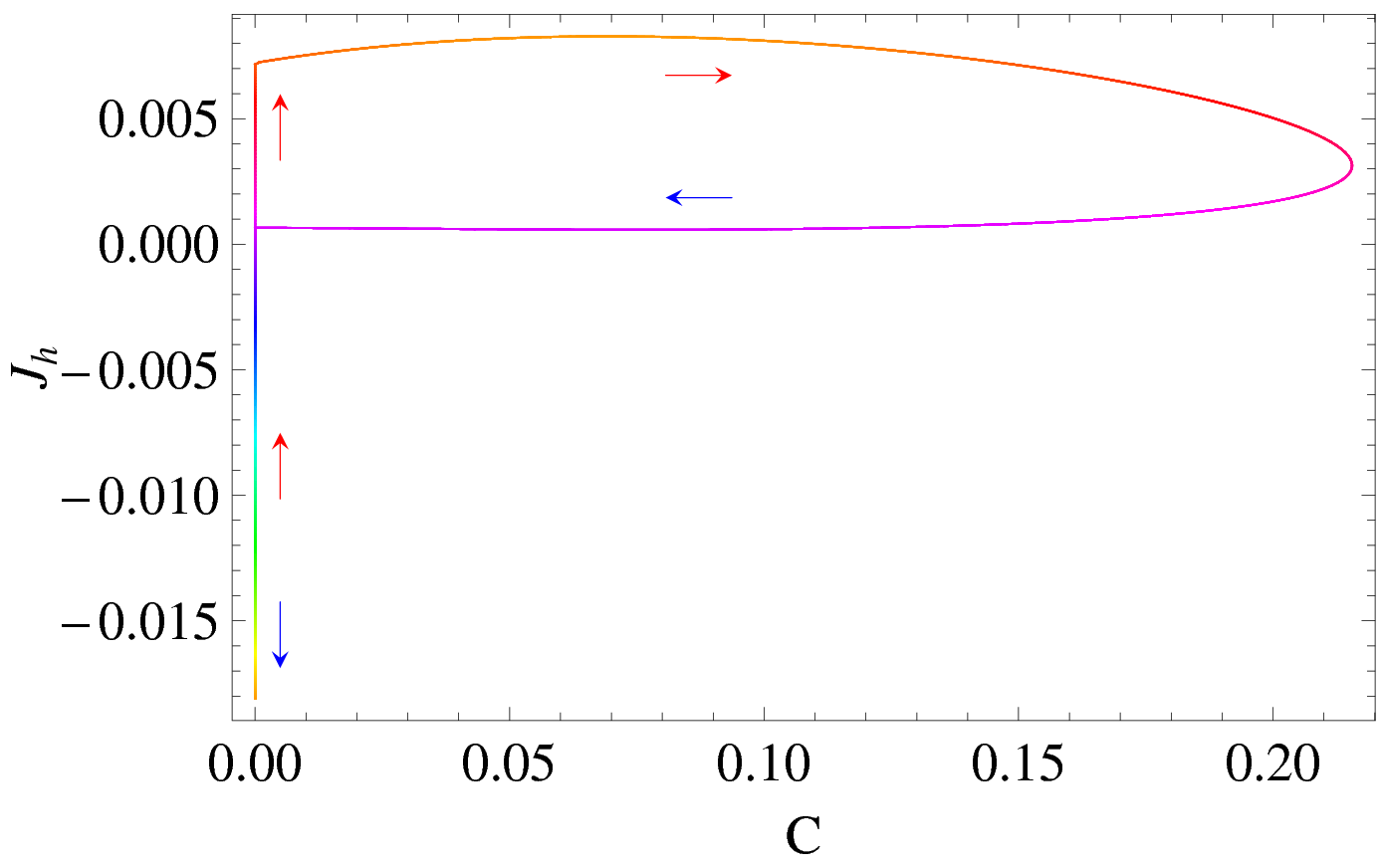}
\end{minipage}
\end{center}
\caption{(Color online) Left: Ground state population $p_{g}$ and
entanglement $C$ {\it vs.} the molecular configuration characterized
by spin-spin distance $d$ for the bosonic heat bath with unit
temperature $T=1$ and the system-bath coupling strength $\kappa=0.01$.
The blue curve is the instantaneous thermal equilibrium state. The arrows
indicate the evolution direction as the molecule oscillates.
Right: Heat current $\mathcal{J}_{h}$ \textit{vs.}
entanglement $C$. The oscillation parameters are
$x_{1}(0)=-x_{2}(0)=-20$, $a=5$, $\tau=100$, and $B_{0}=1.3$,
$B_{1}=2.4$, $\sigma=120$, $J_{0}=1\times 10^{4}$ (see Eq.
(\ref{motion}) and the text thereafter). We use dimensionless parameters
for which the Planck and Boltzmann constants are set to unity, $\hbar=1$, $k_{b}=1$. The temperature $T=1$ means that we set the thermal energy $k_{b}T$ as the unit and express the other values as multiples of the thermal energy. For example,  $B=1$ indicates that $\hbar B/k_{b}T=1$.} \label{QSME}
\end{figure}

Let us first present an intuitive explanation. Consider the following simple process: until time $t=0$, the spins are kept distant (with $J\simeq 0$) and the molecule is in the thermal equilibrium state, with the fraction $p_{g}$ of the
population in the ground state ($|\epsilon_{0}(0)\rangle\simeq
\lvert\downarrow \downarrow\rangle$). If the local level splitting
is sufficiently large such that $\hbar \omega_{0}(0)\gsim k_{B}T$,
$p_{g}$ will be relatively large compared to the other energy
levels. The thermal state will therefore be close to the ground state,
which is, in this case, non-entangled. The adiabatic molecular motion
from the distant configuration to proximity will transform the eigenstates of
$H_{M}(0)$ into those of $H_{M}(t)$; in particular, the ground state
$|\epsilon_{0}(0)\rangle \rightarrow |\epsilon_{0}(t)\rangle$ will
become entangled as the coupling between the spins increases. This
explains, qualitatively, why we may expect entanglement to build up in
{\em one run} of a conformational change, given that the molecular
motion is slow enough to be adiabatic, but at the same time faster
than the thermalization process:
Driven through the classical motion, the system is so-to-speak
``kicked out" of the (separable) thermal equilibrium state,
as can be seen in Fig.~\ref{QSME}.

The above analysis only suggests that entanglement may appear in one
run of a conformational change. However it does not explain how one can
expect to see entanglement on a longer time scale, when the environment
begins to mix the internal states as the molecule continues to oscillate.
It seems that, in the long run, it may (and will) disappear as usual.
What we are interested in, however, is the persistent generation of
dynamic entanglement, thus an extra mechanism is necessary to refresh the state of the molecule by resetting it back to the initial state. It is intriguing that
this role can be played by environmental noise together with oscillatory
motion, both of which naturally exist in biological systems without further
need for control.

Let us now present the numerical results which we obtained by numerically integrating equation \ref{MasterEquation}. We consider the situation where the spin positions are
\begin{equation}
x_{\alpha}(t) =x_{\alpha}(0)+(-1)^{\alpha}a(\cos\frac{2\pi
t}{\tau}-1)
\label{motion}
\end{equation}
where $x_{\alpha}(0)$ are the initial positions, $a$ is the
amplitude of oscillation, and $\tau$ is the oscillation period. For
the local fields we assume Gaussian functions of the spin position
as $B(t)=B_{0}-B_{1}e^{-x^{2}(t)/\sigma }$. For the interaction
between two spins we assume dipole-dipole coupling
$J(t)=J_{0}/d^{3}(t)$ with $d(t)=|x_{1}(t)-x_{2}(t)|$. It can be
seen from Fig. \ref{QSME} that recurrent fresh entanglement appears
on the asymptotic cycle. The thermalizing environment is here
constructive by re-pumping the population into the ground state.

For biological systems, $T\approx 300 K$, the thermal
energy $k_{b}T$ is about $0,025$ eV. Energy scales in bio-molecules
are typically of the order $0.01-0.5$ eV (e.g. for hydrogen bonds or
electronic excitation \cite{Gilmore08,Johnson0708}) and
thus $\hbar \omega_{0}(0)$ can be several times larger than
$k_{b}T$. The system-bath coupling strength $\kappa=0.01$
corresponds to a thermalization time of the order of $\sim  ps$, which
compares e.g. with the timescale of (fast) conformational changes
within bio-molecules and with relaxation times in the FMO complex
\cite{Adolphs06}. These numbers seem to be consistent with our
conclusion that recurrent entanglement might indeed be found in
bio-molecular processes at room temperature.

The oscillating molecule exhibits rather distinct features of
non-equilibrium thermodynamics \cite{Longpaper}, e.g. the entropy
does not always increase with time and reach its saturate value.
The most intriguing feature is a connection between entanglement and
the heat current $\mathcal{J}_{h}$ between the molecule and its
environment, which is defined by the energy dissipated via the heat
bath as \cite{BreuerBook}
\begin{equation}
\mathcal{J}_{h}(t)=Tr\{H_{M}(t) [\mathcal{D} \rho(t) ]\} .
\end{equation}
It can be seen from Fig.~\ref{QSME} that, whenever entanglement
appears, $\mathcal{J}_{h}(t)$ is always positive, i.e. the molecule
tends to absorb heat. This provides some evidence that entanglement
is related with the heat exchange between the molecule
and its environment. Since the oscillating molecule is not in
thermal equilibrium, one cannot adopt the standard definition of
temperature. However, by using the spectral temperature as defined
in \cite{MahlerBook}, one finds that the molecule is effectively
cooled down through the classical motion, even though the attached
thermal bath is always at a fixed higher temperature. This
interesting observation is consistent with our common intuition
that entanglement appears as the system is cooled down.

\begin{figure}[ht]
\begin{center}
\begin{minipage}{8.8cm}
\includegraphics[width=4.3cm]{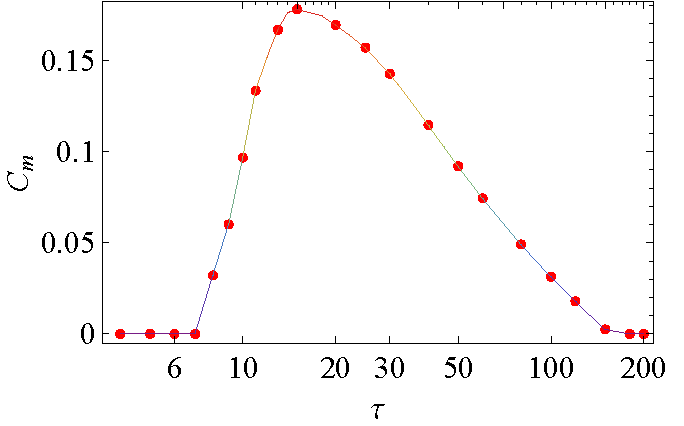}
\hspace{0cm}
\includegraphics[width=4.3cm]{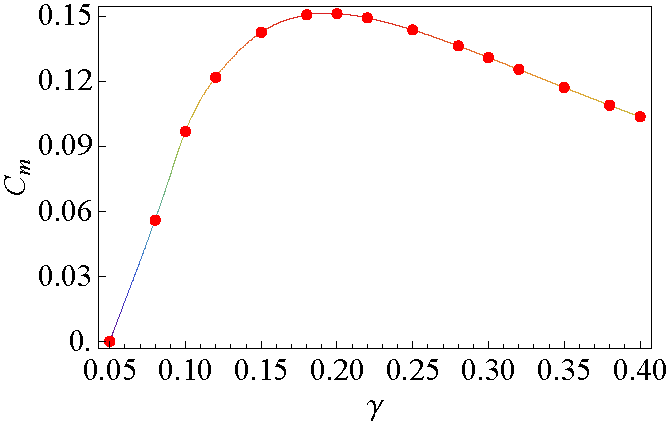}
\end{minipage}
\caption{(Color online) Left: The maximal value of entanglement on
the asymptotic cycle $C_{m}$ {\it vs.} the oscillation period
$\tau$ (logarithmic coordinate) with $\gamma=0.1$. Right: $C_{m}$
{\it vs.} $\gamma$ with $\tau=10$. The other parameters are $s=0.2$,
$x_{1}(0)=-x_{2}(0)=-25$, $a=10$, and $B_{0}=B_{1}=1.2$,
$\sigma=120$, $J_{0}=1.2\times10^{3}$.} \label{CMT}
\end{center}
\end{figure}

The results discussed so far have been obtained by modeling the
noisy environment as an Ohmic bosonic heat bath.  Clearly, this is a
highly idealized model and in any real biological scenario we have
far more complicated interactions, e.g. with the surrounding
hydration shell and the bulk solvent \cite{Fenimore04,Gilmore08}.
We found similar results also with other decoherence models, based on collision-type interactions of the environment with the system \cite{Longpaper}. These can be described by Lindblad generators $L^{(\alpha)}_{g}=\sqrt{\gamma s} \sigma^{(\alpha)}_{+} \quad \text{and}
 \quad  L^{(\alpha)}_{d}=\sqrt{\gamma (1-s)} \sigma^{(\alpha)}_{-}$,
where $\gamma$ is the collision-induced effective relaxation rate and $s$ is the mean excitation of a spin in thermal equilibrium. In Fig.~\ref{CMT} we demonstrate the competition between the constructive and destructive effects of environmental noise in such a model, which yields an optimal value for the oscillation period $\tau$ to establish entanglement, see Fig.~\ref{CMT}a.
For short oscillation periods, efficient thermalization becomes more important: with growing rate $\gamma $, the increase in reset efficiency more than compensates the decrease in efficiency of entanglement generation, see Fig.~\ref{CMT}b. In this regime, the net effect of environmental noise is constructive.
The dependence of the entanglement on $\tau$ and $\gamma$ seems reminiscent of what happens in the very nice example of quantum ``stochastic resonance",
which has been described in a quantum-optical context \cite{Plenio02}. That phenomenon is, however, fundamentally different from ours, as it involves a bath at zero temperature, and the noise is the main driving force.

Given the complexity of biological systems, how to characterize
biological environment is far from been understood. The effect we
presented --- i.e. the existence of persistent entanglement --- is
to a very large extent independent of the precise details of the
classical movement, and the thermal bath. Of course, the detailed
characteristics of the entanglement (how much entanglement, how does
it vary with time, etc.) depend on the driving motion and the
specific environment, but the very existence of persistent
entanglement is generic (extensible to different types of classical
motions, spectral density of thermal bath and also to the
non-Markovian environment). A number of different models illustrating this generic property are presented in the Supplementary Information.


One can test the feasibility of our
model by simulating an oscillating molecule in a noisy environment.
Two internal levels of trapped ions can encode an effective two-level
system. The system Hamiltonian in the form of Eq.(\ref{Hamiltonian})
is implementable via state-dependent optical dipole forces \cite{Porras04}.
The classical oscillation, the essence of which is to introduce a time-dependent
Hamiltonian, can be simulated by tuning the interaction strength and
the transverse fields, which is achievable e.g.\ by changing the
amplitudes of laser beams \cite{Friedenauer08,Wineland0805,Wunderlich}.
Decoherence can be simulated by e.g. applying random pulses to induce
different decoherence channels. One can also simulate the bosonic bath
by engineering the coupling between ions and vacuum modes of the
electromagnetic field through laser radiation \cite{Poyatos96}.
Entanglement can finally be detected by performing quantum state
tomography as in \cite{RoosPRL}. Other implementations are conceivable e.g.
using quantum dots mounted on the tips of oscillating cantilevers
\cite{Bouwmeester}.

\section{Supplementary Information}

This is supporting material for our paper. We present the essential
steps in the derivation of the quantum master equation for the
oscillating molecule in contact with bosonic heat baths, and the
calculation of the concurrence which quantifies the entanglement
between the spins. We also describe the spin-gas model as an
alternative model for the environment and show that it gives rise to
qualitative similar features of reset and recurrent entanglement.
Finally, we analyze the competing features of environmental noise
for the generation and degradation of entanglement in such a model.

{\bf Methods.--} To account for the effect of the environment on the
oscillating two-spin molecule, we have studied different models, two
of which we discuss here. In the first model - the bosonic heat bath
- each qubit is coupled to an independent thermal bath of harmonic
oscillators. This is a well-known decoherence model and the
derivation of the master equation (L2) \footnote{References to
equations and figures in the Letter start with an ``L", i.e. ``eqn.
(L2)" means ``eqn. (2) in the Letter".} follows standard techniques
(see e.g. \cite{BreuerBookSI}), with the important difference that
in our case the Hamiltonian $H_{M}(t)$ of the system is \emph{time
dependent}, i.e. its instantaneous energy spectrum changes as the
qubits move. This makes the analysis much more complicated, but
under certain conditions it can be still be described by an equation
of type (L2), albeit with time-dependent Lindblad operators.

In the second model - the spin-gas model \cite{Hartmann05SI} - each
qubit is subject to random, collision-type interactions with a
``background gas'' of other spin particles. These processes can lead
to both local spin exchange and de-phasing. Here we model these
processes again by a Lindblad-type master equation
\cite{Hartmann07SI}, but we mention that a numerical treatment
including memory effects in the environment, which gives rise to
non-Markovian de-coherence, can be given \cite{Hartmann05SI}.

We calculate the entanglement that is generated during the molecular
motion, using the two-qubit measure of concurrence
\cite{Wootters98SI}.

\vspace{12pt} {\bf Master equation for the oscillating molecule in
contact with bosonic heat baths.--} The derivation of the master
equation follows standard arguments used in reservoir theory (see
e.g. \cite{BreuerBookSI}), but with some important modifications due
to the time dependence of the problem.

One should point out from the outset that the derivation of the
master equation rests on a series of assumptions (including e.g. the
Born and Markov approximation and the secular or rotating-wave
approximation), neither of which we expect to be very well satisfied
in real biological systems.

What the master equation {\em does provide}, however, is a dynamical
process that exhibits the essential features that we expect to be
most relevant in our system of consideration: A process that is {\em
disentangling} and that leads to de-coherence and thermalization in
the subsystem, the two-spin molecule. Whether or not these processes
follow, e.g., an exponential decay is not so essential for the main
argument.

The total Hamiltonian of the system and the environment can be
written in the form
\begin{equation}
H_{tot}(t)=H_{M}(t)+H_{B}+H_{MB}
\end{equation}
where $H_{M}(t)$ is the time-dependent Hamiltonian (L1) of the
two-spin molecule and $H_{B}$, $H_{MB}$ describe the oscillator bath
and the molecule-bath interactions, respectively.  We assume
dissipative coupling, in which case the latter can be written in the
form
\begin{equation}
H_{MB} = \sum_{\alpha}\sigma_{x}^{( \alpha )}\otimes B_{\alpha}
\label{HMB}
\end{equation}
where $B_{\alpha}=B_{\alpha}^{\dag}$ denote the collective bath
degrees of freedom that couple to the $\alpha$th spin. The
interaction constants have been absorbed in the $B_{\alpha}$s.

Within the usual Born-Markov approximation, one obtains an equation
of motion for the two-spin molecule which, in the interaction
picture, has the form
\begin{equation}
\frac{\partial}{\partial t}\rho (t) = -\int_{0}^{t}ds\;
Tr_{B}[H_{MB}(t),[H_{MB}(t-s),\rho (t)\otimes \rho_{B}]],
\label{BME}
\end{equation}
where $\rho (t) = Tr_{B}\rho _{MB}(t)$ represents the reduced
density matrix of the two-qubit molecule after tracing out the
degrees of freedom the thermal bathes.

To perform the secular or rotating wave approximation, we expand the
spin operators $\sigma_{x}^{( \alpha )}$ in (\ref{HMB}) into the
basis of instantaneous eigenstates $|\epsilon_{i}(t)\rangle$ with
eigenvalues $\epsilon_{i}(t)$ ($i=0,..,3$) of the system Hamiltonian
$H_{M}(t)$, i.e.\
\begin{eqnarray}
\sigma_{x}^{(\alpha)} &=&\sum_{\omega (t)} S_{\omega (t)}^{(\alpha)} \\
&=&\sum_{\omega (t)}\sum_{\epsilon_{i}(t)-\epsilon_{j}(t)=\omega
(t)} S_{ij}^{(\alpha )}(t) |\epsilon_{i}(t)\rangle \langle
\epsilon_{j}(t)|\,.
\end{eqnarray}
In the first line, the operators $S_{\omega (t)}^{(\alpha)}$
describe intra-molecular transitions with frequency $\omega (t)$ and
the summation runs over all resonant transition frequencies; in the
second line, $S_{ij}^{(\alpha )}(t)=\langle \epsilon _{i}(t)|\sigma
_{x}^{(\alpha)}|\epsilon_{j}(t)\rangle $ are the corresponding
transition matrix elements and the summation runs over all states
with matching energy eigenvalues $\epsilon_{i}(t)$,
$\epsilon_{j}(t)$. In the interaction picture, which is used in the
derivation of the master equation, we then write
\begin{equation}
\sigma_{x}^{(\alpha )}(t)=\sum_{\omega (t)}U^{\dagger }(t)S_{\omega
(t)}^{(\alpha)}U(t)
\end{equation}
where $U(t)$ is the unitary evolution generated by the system
Hamiltonian, $U(t)=\mathcal{T}e^{-i\int_{0}^{t}H_{M}(s)ds}$, and
$\mathcal{T}$ denotes the time ordering operator.

In a situation where the system's evolution is \emph{adiabatic},
i.e. slow enough to avoid energy-changing transitions, the
eigenstates will only pick up a dynamic phase (under the coherent
evolution of the time-dependent system Hamiltonian)
\cite{LongpaperSI}. The time dependence of the spin operators in the
interaction picture then acquires the simple form
\begin{equation}
\sigma_{x}^{(\alpha )}(t)=\sum_{\omega
(\cdot)}e^{-i\int_{0}^{t}\omega
(s)ds}\sum_{\epsilon_{i}(t)-\epsilon_{j}(t)=\omega (t)}
S_{ij}^{(\alpha )}(t)|\epsilon _{i}(0)\rangle \langle \epsilon
_{j}(0)|.
\end{equation}
Upon inserting this into (\ref{BME}) and applying (a generalization
of) the secular approximation \cite{BreuerBookSI}, we obtain a
master equation of the form (L2) with implicitly time-dependent
Lindblad operators \cite{LongpaperSI}. The properties of the baths
thereby enter through the following Fourier-type transform of the
bath correlation functions
\begin{equation}
\Gamma_{\alpha\beta}[\omega(\cdot),t]=\int_{0}^{\infty}ds
e^{i\int_{t-s}^{t}\omega (s)d s}\langle
B_{\alpha}^{\dagger}(s)B_{\beta}(0)\rangle \label{FBB}
\end{equation}
which is different from the exact Fourier transform obtained in case
of a time-independent system Hamiltonian. For independent baths, we
only get a contribution for $\alpha=\beta$. The Markov approximation
assumes that the correlation functions $\langle
B_{\alpha}^{\dagger}(s)B_{\alpha}(0)\rangle$ decay fast compared to
the relaxation time, which means that their real and imaginary part
can essentially be replaced by the delta function $\delta (t)$ and
its time derivative $\delta^{\prime}(t)$, respectively. It is
therefore consistent to apply, for small values of $s$, the
following approximation for the integral in (\ref{FBB})
\begin{equation}
\int_{t-s}^{t}\omega (t^{\prime})dt^{\prime} \simeq \omega(t) s
\end{equation}

In summary, we can thus write
\begin{equation}
\Gamma_{\alpha\beta}[\omega(\cdot),t]\simeq
\delta_{\alpha\beta}\Gamma_{\omega(t)}
\end{equation}
where the function $\Gamma_{\omega(t)}$ now depends only on the
value of the frequency $\omega(t)$ at the time $t$. Upon
transforming back to the Schr{\"o}dinger picture, $
|\epsilon_{i}(0)\rangle$ in the transition operators
$\sigma_{x}^{(\alpha )}(t)$ are mapped to $|\epsilon_{i}(t)\rangle$,
and we finally obtain a master equation with the implicitly
time-dependent Lindblad generators
\begin{equation}
L_{\mu}\equiv
L_{\alpha}(\omega(t))=\Gamma^{1/2}_{\omega(t)}\sum_{\Delta_{i
j}(t)=\omega(t)} S^{(\alpha)}_{ij}(t) |\epsilon_{i}(t)\rangle
\langle \epsilon_{j}(t)|.
\end{equation}
The index $\mu \equiv \{\alpha, \omega(t)\}$ in Eq. (L2) of the
Letter runs here over $\alpha =1,2$ and over the allowed values of
$\omega(t)$. The relevant quantity of the heat bath which enters
$\Gamma_{\omega(t)}$ is its spectral density function. If we assume
an Ohmic spectral density with infinite cut-off frequency, we obtain
\begin{equation}
\Gamma_{\omega(t)}=\kappa\omega(t)(1+ N_{\omega(t),\beta})
\label{OSD}
\end{equation}
where $N_{\omega(t),\beta}$ is the bosonic distribution function at
inverse temperature $\beta$, i.e.,
$N_{\omega(t),\beta}=1/(e^{\omega(t)\beta}-1)$. The master equation
(L2) with these Lindblad operators was used to calculate the data
shown in Fig. L3.

\vspace{12pt} {\bf Concurrence and static thermal entanglement.--}
To measure the dynamic entanglement generated during molecular
oscillation, we compare it with the thermal equilibrium state when
the molecular configuration is fixed at any distance. In such a
``static configuration", both the Hamiltonian and the Lindblad
operators are time independent. Furthermore, the derived quantum
master equation is then {\em mixing} and the molecule will always be
driven to its thermal equilibrium state \cite{BreuerBookSI}
corresponding to the reservoir temperature $\beta =1/T$. For each
specific molecular configuration, with fixed spin-spin interaction
strength $J$ and local electric or magnetic fields $B$, the thermal
equilibrium state reads
\begin{equation}
\rho _{th}=e^{-\beta H_{M}}/\mathcal{Z}=\frac{1}{\mathcal{Z}}\left(
\begin{array}{cccc}
r_{00} & 0 & 0 & r_{03} \\
0 & r_{11} & r_{12} & 0 \\
0 & r_{21} & r_{22} & 0 \\
r_{30} & 0 & 0 & r_{33}%
\end{array}%
\right)   \label{TES}
\end{equation}%
where $\mathcal{Z}=Tr(e^{-\beta H_{M}})$ is the partition function
and the matrix representation refers to the computational product
basis. The non-zero entries of the above matrix are given by
\begin{eqnarray}
r_{00} &=&e^{\mathcal{E}\beta }-2\sinh {(\mathcal{E}\beta )}/(1+\eta
^{2}),\qquad  \\
r_{11} &=&r_{22}=\cosh {(J\beta ),} \\
r_{33} &=&e^{-\mathcal{E}\beta }+2\sinh {(\mathcal{E}\beta
)}/(1+\eta
^{2}),\quad  \\
r_{03} &=&-J\sinh {(\mathcal{E}\beta )}/\mathcal{E},\quad r_{12}=-\sinh {%
(J\beta )}
\end{eqnarray}%
where $\mathcal{Z}=2[\cosh {(\mathcal{E}\beta )}+\cosh {(J\beta
)}]$, $\mathcal{E}=(4B^{2}+J^{2})^{1/2}$, and $\eta
=(\mathcal{E}-2B)/J$. To quantify the two-qubit entanglement, there
exist various kinds of entanglement measurements. We choose the
concurrence $C$ \cite{Wootters98SI}, which is defined as $C=\max
\left\{ 0,\lambda _{1}-\lambda _{2}-\lambda _{3}-\lambda
_{4}\right\} $, where the $\lambda _{i}$s are the square roots of
the eigenvalues of $\rho\tilde{\rho}$ in decreasing order
\cite{Wootters98SI}, with  $\tilde{\rho}=(\sigma _{y}\otimes \sigma
_{y})\rho ^{\ast}(\sigma _{y}\otimes \sigma _{y})$. For the thermal
equilibrium state (\ref{TES}), one obtains
\begin{equation}
C(\rho _{th})=\frac{2}{\mathcal{Z}}\max {%
\{0,|r_{12}|-(r_{00}r_{33})^{1/2},|r_{03}|-(r_{11}r_{22})^{1/2}\}}.
\end{equation}
Using the explicit expressions for $r_{12},r_{00}, r_{33}$ one finds
that $|r_{12}|-(r_{00}r_{33})^{1/2}\leq 0$ and
$|r_{03}|-(r_{11}r_{2})^{1/2}=
\frac{|J|}{\mathcal{E}}\sinh{(\mathcal{E}\beta )}-\cosh {(J\beta
)}$.

The static thermal entanglement can thus be written as
\begin{equation}
C(\rho _{th})=\frac{2}{\mathcal{Z}}\max {\{0,\frac{|J|}{\mathcal{E}}\sinh {(%
\mathcal{E}\beta )}-\cosh {(J\beta )}\}}.
\end{equation}

In order to illustrate how the static entanglement changes as the
temperature increases, we calculate the first derivative of $C$ with
respect to $\beta$. After some straightforward calculations, it can
be seen that
\begin{equation}
\partial C(\rho_{th})/\partial \beta \geq 0
\end{equation}
which means that the static entanglement always decreases as the
temperature increases. This is consistent with our intuition that
there exists a critical temperature $T_{c}$, above which no static
entanglement can survive. In other words, for any fixed molecular
configuration, entanglement will eventually vanish when the
reservoir is too hot. In our Letter, we consider exactly such a
situation. The temperature of the environment is so high that the
thermal state is separable (non-entangled) at all possible molecular
configurations, i.e. $T > \max\{T_{c}\}$.

\vspace{12pt} {\bf Spin-gas model for the environment.--} The
environment of bio-molecular systems is rather complex
\cite{Frauenfelder09SI} and not yet fully understood, and some of
its features may not be well-described by a thermal bath model of
harmonic oscillators. To check whether the observed effects are
robust, we have also considered an alternative model -- the
so-called spin gas model \cite{Hartmann05SI,Hartmann07SI} -- for the
environment. In this model, we assume collisions between the
molecular spins and other, randomly moving spin-particles that
constitute the environment. The collisions induce local energy
dissipation (i.e. spin exchange) and de-phasing, which leads to
de-coherence and, if left alone, quickly destroys all entanglement
between the molecular spins.\footnote{The spin-gas model should not
be confused with the spin-\emph{bath} model \cite{Stamp00SI}, which
is similar but assumes a static distribution of random couplings
between the molecular and the environmental spins.}

\begin{figure}[htb]
\begin{center}
\begin{minipage}{8.5cm}
\includegraphics[width=4.2cm]{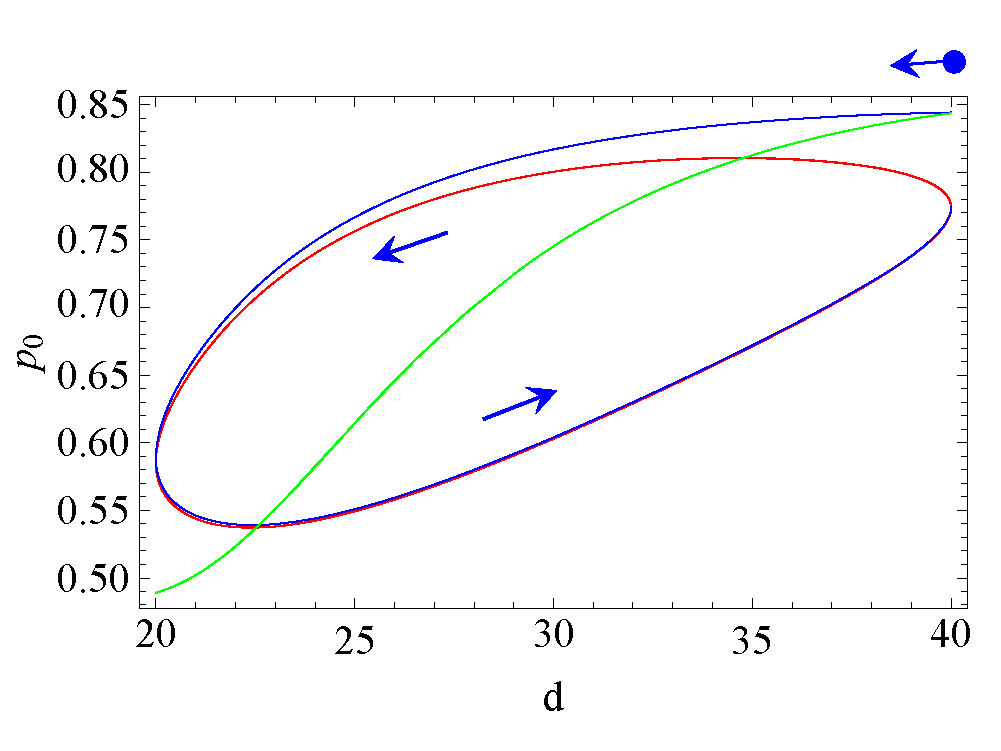}
\includegraphics[width=4.2cm]{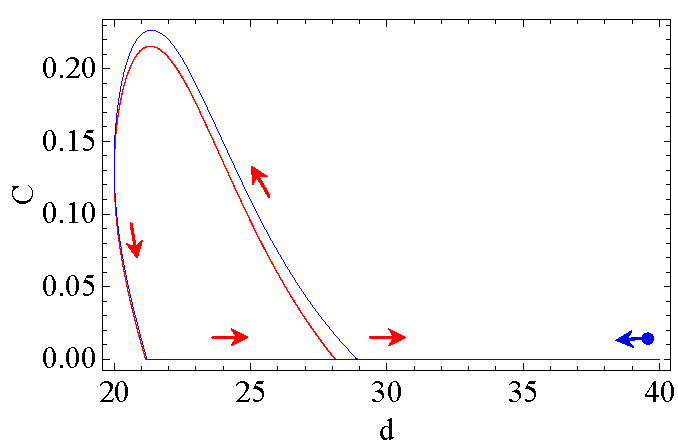}
\end{minipage}
\begin{minipage}{8.7cm}
\includegraphics[width=4.2cm]{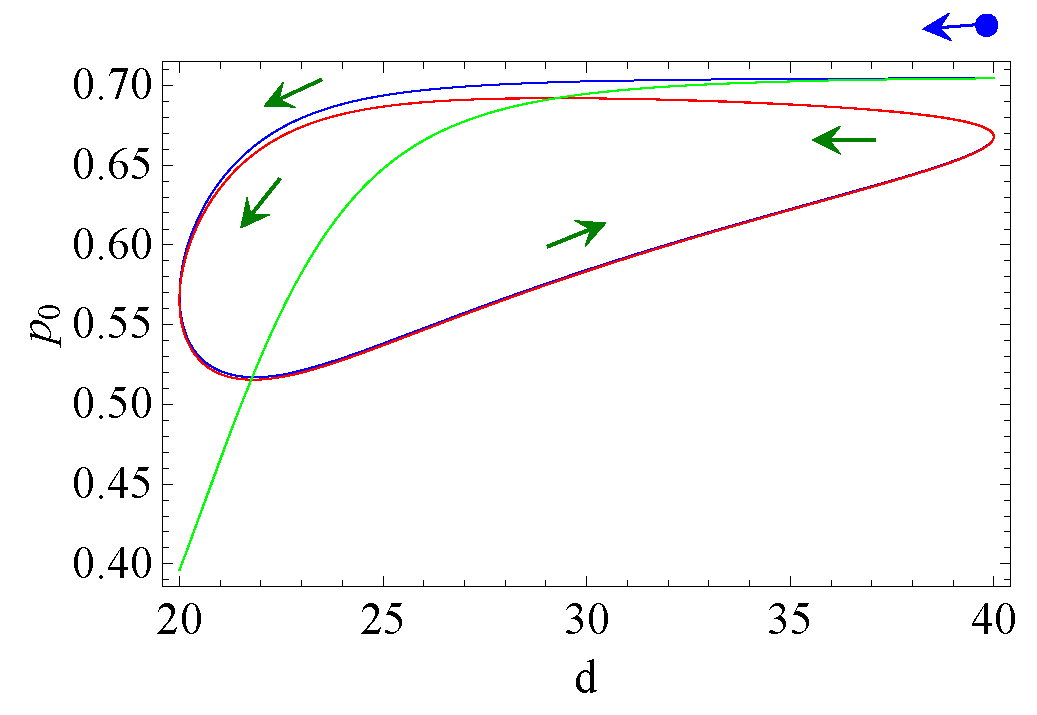}
\includegraphics[width=4.4cm]{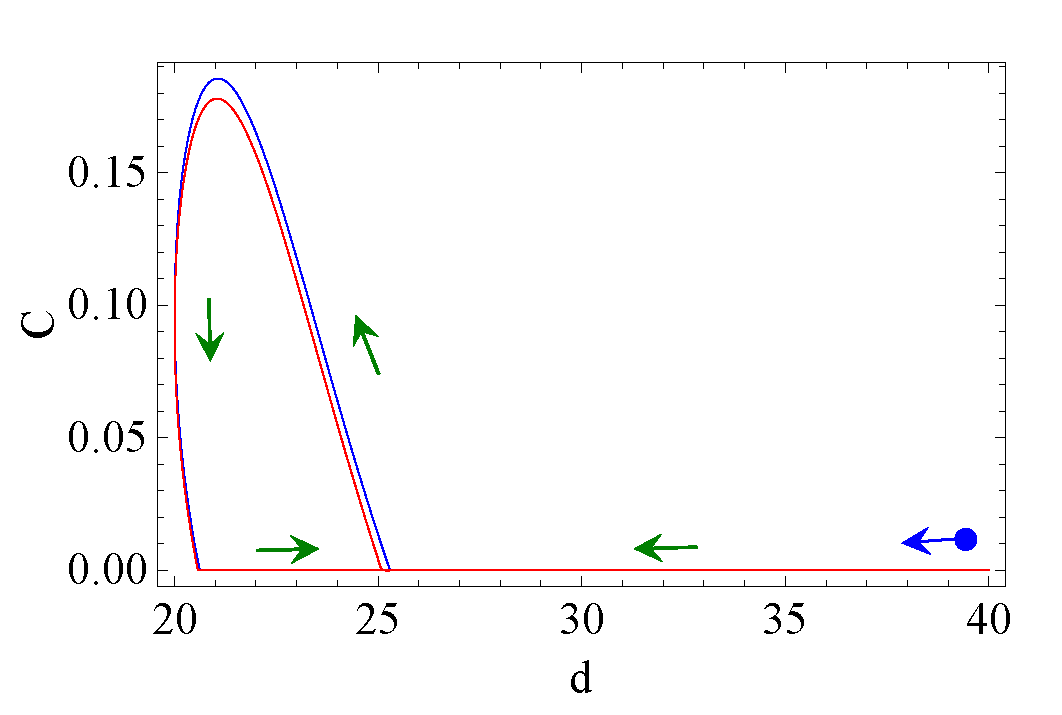}
\end{minipage}
\end{center}
\caption{Ground state population $p_{0}$ (left) and dynamic
entanglement $C$ (right) \textit{vs.} the molecular configuration
$d$. Upper: Bosonic thermal bath at temperature $T=1$ and
$\protect\kappa =0.01$; Lower: Spin gas model with $s=0.16$ and
$\protect\gamma=0.025$. The oscillation parameters are the same as
in Fig.~L3 of our Letter. The first cycle and the asymptotic cycle
are indicated by the blue and red curves respectively. The green
curves (left) represent the instantaneous steady state corresponding
to each molecular configuration. The blue dot and arrows indicate
the starting point and the evolution direction.} \label{CME}
\end{figure}

The spin-gas model of the environment has been described in
Refs.~\cite{Hartmann05SI,Hartmann07SI}. For Ising-type interactions,
one can calculate the time evolution of small subsystems exactly,
for environments consisting of up to $10^{5}-10^{6}$ particles,
without any approximations. Under certain conditions, one can again
derive a master equation by considering the effect of random
collisions on a coarse-grained time scale \cite{Hartmann07SI}. For
the present purpose, we will employ such a phenomenological
description, but we emphasize that non-markovian and collective
effects in the environment can be taken into account
\cite{Hartmann05SI}.

The effect of random collisions leads to Lindblad generators of the
form $L_{i} \in \{L^{(\alpha)}_{g},L^{(\alpha)}_{d}|\alpha=1,2\}$
with
\begin{equation}
L^{(\alpha)}_{g}=\sqrt{\gamma s} \sigma^{(\alpha)}_{+} \quad
\mbox{and} \quad L^{(\alpha)}_{d}=\sqrt{\gamma (1-s)}
\sigma^{(\alpha)}_{-}
\end{equation}
where $\sigma_{\pm}^{(\alpha)}=(\sigma_{x}^{(\alpha)}\pm i
\sigma_{y}^{(\alpha)})/2$ are the Pauli ladder operators for a
two-level system. The resulting master equation \cite{Hartmann07SI}
describes local energy gain and loss processes (``spin exchange")
with the effective rate $\gamma >0 $, while $s$ is related to the
temperature and determines the equilibrium distribution of the local
excitation \cite{Hartmann07SI}. Without loss of generality, we may
assume that the $\gamma >0 $ and $0 \leq s\leq 1/2$. If $s$ is
larger than a critical value $s_{c}$, no static entanglement can
exist \cite{LongpaperSI}, similar as in the case of the bosonic heat
bath in the previous section.

Even though this model is quite different from the bosonic heat
bath, it has certain features in common, for example, it is
disentangling and mixing. Remarkably, we find the same phenomenon as
in case of the bosonic heat bath, namely a persistent recurrence of
fresh entanglement in a regime where the static entanglement
vanishes for all molecular configurations.

In Fig.~\ref{CME} we compare the evolution of the oscillating
molecule for the spin gas model and the bosonic heat bath model. The
upper panel of Fig.~\ref{CME} reproduces the left 3D plot in Fig.~L3
(left) of our Letter by projecting it onto two dimensional curves.
The lower panel shows the same evolution for the spin gas. It can be
seen that the \emph{qualitative} features are robust: During the
first oscillation, entanglement builds up when the spins approach
each other, while the evolution subsequently converges towards an
asymptotic cycle on which the entanglement periodically recurs. This
observation strengthens our claim that this feature is robust and
does not seem to depend on the detailed features of the environment.

\vspace{12pt} {\bf Competing effects of environmental noise on
dynamic entanglement.--} Another benefit of the phenomenological
spin-gas model is that it allows one to enter the regime of short
oscillation periods and to clearly illustrate the competition
between the constructive and the destructive effects of the
environmental noise.\footnote{We remark that the task of deriving a
simple master equation that is also valid for fast oscillations
(i.e. beyond the adiabatic approximation) becomes very hard in the
case of the bosonic thermal bath.} In Fig.~L4, we have plotted the
maximal value of entanglement that is assumed on the asymptotic
cycle, in the case where $s=0.2 > s_{c}$ (i.e. the equilibrium state
is separable for all possible molecular configurations). The left
plot displays the maximally achievable entanglement for different
oscillation periods. It can be seen that the occurrence of dynamic
entanglement strongly depends on the oscillation period; there are
competing effects of the environmental noise which give rise to an
optimal oscillation period where the effect dynamic entanglement is
most pronounced.

\begin{figure}[htb]
\begin{center}
\hspace{-0.8cm}
\includegraphics[width=9cm]{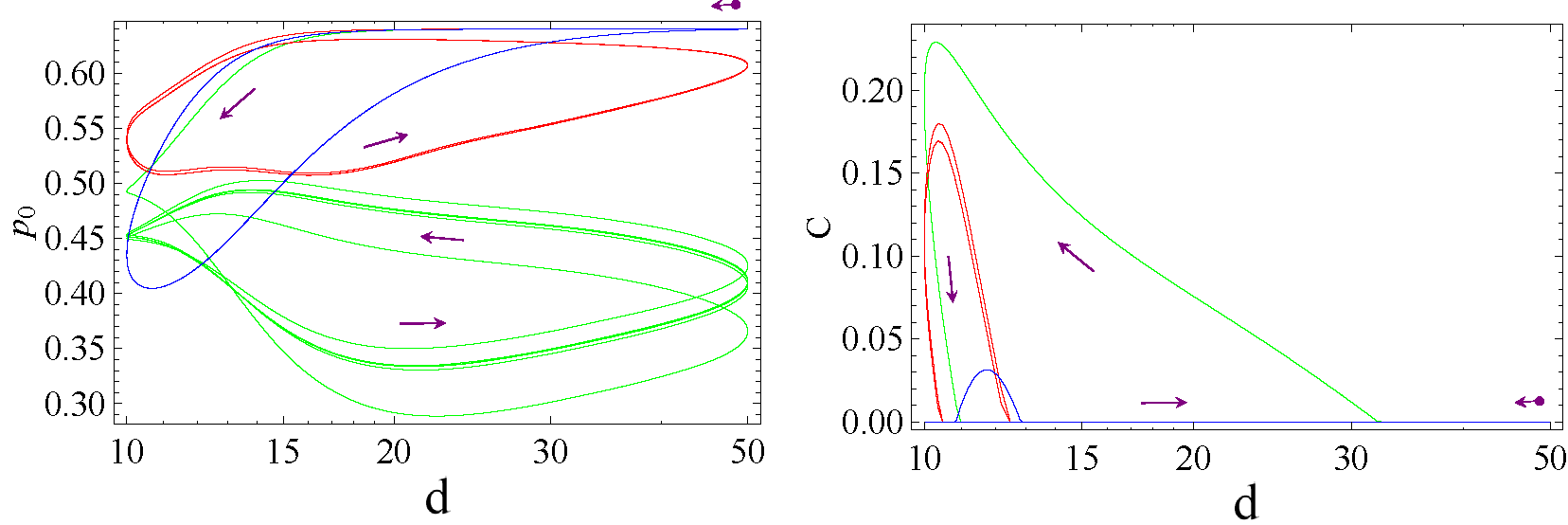}
\end{center}
\caption{Ground state population $p_{0}$ (left) and dynamic
entanglement $C$ (right) \textit{vs.} the molecular configuration
$d$ (logarithmic coordinate). The oscillation periods are
$\protect\tau=6$ (green), $20$ (red) and $100$ (blue) respectively.
The dot and arrow indicate the starting point and the evolution
direction. Right: The green curve (with positive values) only
represents the first cycle, while subsequent cycles collapse onto
the abscissa, without entanglement.  The other parameters are the
same as in Fig.~L4 (left) of our Letter.} \label{LMS}
\end{figure}

To understand the results in Figs. L4 and \ref{LMS} better, it is
worth looking at the detailed time evolution of the entanglement and
the ground state population for three typical oscillation periods.

First, consider a very long oscillation period, e.g. $\tau=100$.
Here,the molecule is almost completely reset by thermalization to
equilibrium when the spins are spatially separated (distant
configuration), with a large ground state population of up to $\sim
65\%$, see Fig.~\ref{LMS} (blue curve). Since in this regime the
coherent evolution of the molecule is adiabatic, the population of
the instantaneous eigenstates of the system Hamiltonian remains
approximately constant, while the off-diagonal elements remain
negligible. When two spins come closer, they start interacting and
the ground state becomes entangled (entanglement generation regime).
If there were no dissipation, the high population of the ground
state alone would be sufficient to generate entanglement. However,
while the spins approach each other, the energy separation between
the lowest lying levels decreases and the dissipation starts
re-populating the levels. This drives the molecular state into the
separable regime and diminishes its entanglement, with only little
entanglement surviving. For a moderate oscillation period, e.g. $
\tau=20$ (red curve), the dissipation still has enough time to reset
the system while it passes through the distant configuration, with a
ground state population similar as for as $\tau=100$. In the
entanglement generation regime, however, the destructive effect of
the dissipation is now much smaller than for long oscillation
period, which leads overall to more entanglement. Finally, for a
very short oscillation period, e.g. $\tau=6$, the destructive effect
during the entanglement generation regime is even smaller. On the
other hand, the reset effect in the distant configuration is greatly
suppressed, since the system does not have enough time to thermalize
and to repopulate the ground state. Thus, even though the transient
entanglement is larger in the first period, as expected, it will
diminish in subsequent runs and cannot be sustained on the
asymptotic cycle, due to the lack of an effective reset mechanism.

\vspace{12pt} {\bf Generic features of persistent dynamic
entanglement in noisy environment.---} In the main text, we have
used a simple model, namely a harmonically oscillating molecule in
an Ohmic thermal bath, to demonstrate the essential mechanism for
persistent dynamical entanglement to occur in a de-coherent
environment where no static entanglement can exist. Here we show
that neither the harmonic oscillatory motion nor the Ohmic thermal
bath are indispensable for such an effect. The existence of
persistent dynamic entanglement is to a very large extent
independent of the precise details of the classical motion and
thermal environment.

\begin{figure}[htb]
\begin{center}
\begin{minipage}{6.5cm}
\includegraphics[width=6.1cm]{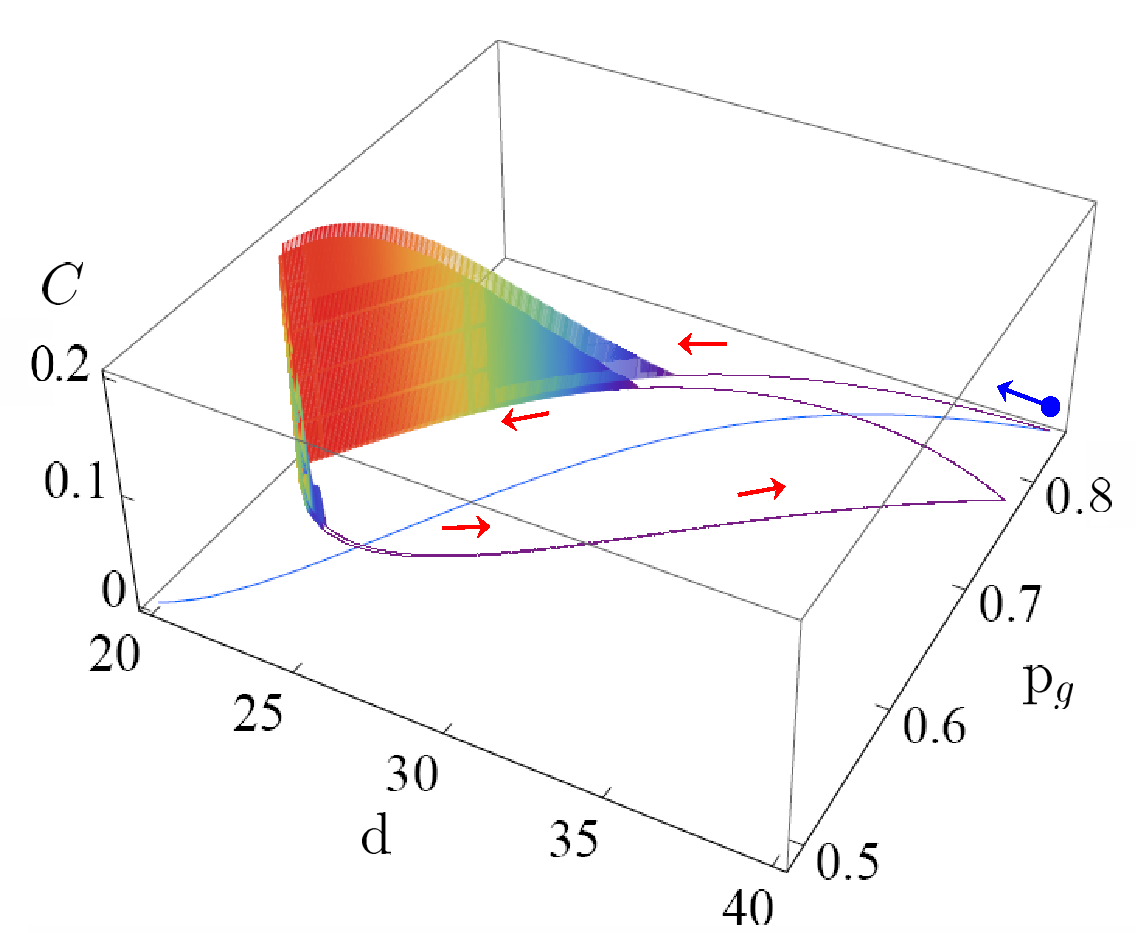}
\end{minipage}
\end{center}
\caption{Ground state population $p_{g}$ and entanglement $C$ {\it
vs.} the molecular configuration characterized by spin-spin distance
$d$ for the bosonic heat bath. The spins move towards (away) from
each other with a constant speed. The other parameters are the same
as Fig. L3.} \label{ConSpeed}
\end{figure}

In Fig.~\ref{ConSpeed}, we consider a model where the spins move
towards (away) from each other with a constant speed, and observe
similar results as for the harmonic oscillatory motion. The same
effect can also be seen in a scenario of stochastic movements
\cite{LongpaperSI}. In short, the detailed characteristics of the
entanglement (how much entanglement, how does it vary with time,
etc.) depend on the driving oscillation, but the very existence of
persistent entanglement is generic. All that is needed is that the
classical motion obeys two conditions: (i) is adiabatically slow but
sufficiently fast compared to de-coherence and (ii) it spends long
enough time at the far end for thermalization to occur.

\begin{figure}[htb]
\begin{center}
\begin{minipage}{9cm}
\includegraphics[width=4.2cm]{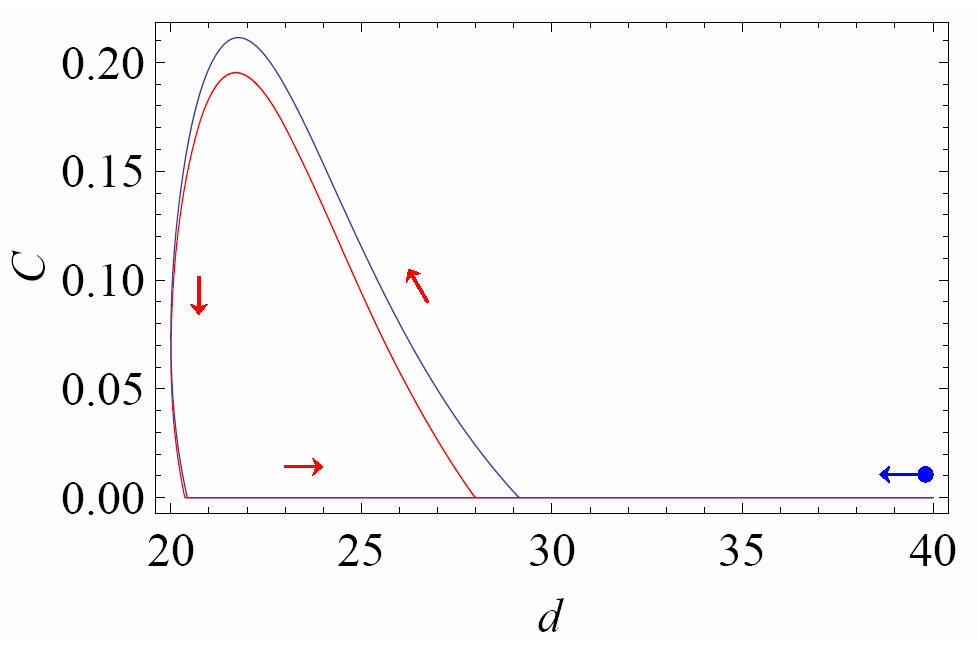}
\includegraphics[width=4.2cm]{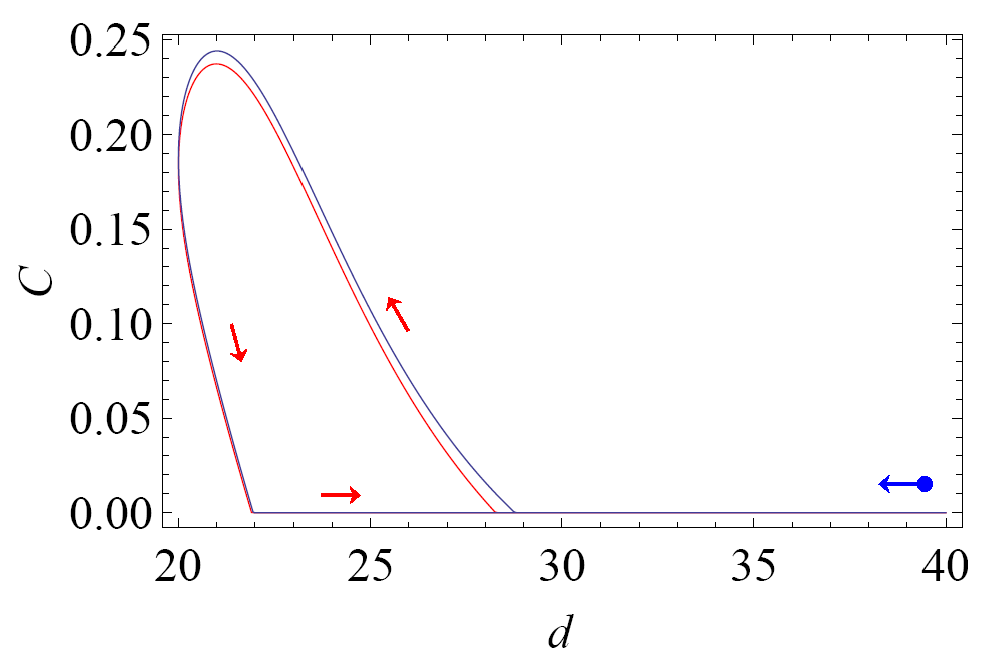}
\end{minipage}
\end{center}
\caption{Dynamic entanglement $C$ \textit{vs.} the molecular
configuration $d$ for the sub-ohmic (left, $s=0.8$) and supra-ohmic
(right, $s=1.2$) thermal bath. The other parameters are the same as
Fig. L3. The blue dot and red arrows indicate the starting point and
the evolution direction. The first cycle and the asymptotic cycle
are indicated by the blue and red curves, respectively.}
\label{SubSupOhmic}
\end{figure}

Regards the model for the bath, we have so-far used the Ohmic bath
(as well as the spin-gas model) as a example. However, our results
are also valid for other forms of spectral densities. It can be seen
from Fig.~\ref{SubSupOhmic} that the effect of persistent dynamic
entanglement is not restricted to the simple Ohmic bath, but occurs
also for the sub-Ohmic and supra-Ohmic bath, i.e. for a spectral
density $\sim \omega^{s}$ with $s<1$ or $s>1$. We have also found
that the present mechanism works even with the spectral density from
the solvent and protein environment \cite{McK08SI}, e.g. a spectral
density in the form of $\sim \frac{1}{\omega}$. Finally, by using
the numerical method of quasi-adiabatic propagator path integral
\cite{Mark95aSI,Mark95bSI}, we have extended the results to the
non-Markovian environment with finite memory time, and still see the
generic effect discussed in the main text. More details will be
presented in \cite{LongpaperSI}.



\begin{thebibliography}{99}

\bibitem{Abbott08}
Abbott, D., Davies, P. C. W. \& Pati, A. K. (eds.)
\newblock \emph{Quantum Aspects of Life} (World Scientific, Singapore, 2008).

\bibitem{Briegel0806}
Briegel, H.~J. \& Popescu, S.
\newblock Entanglement and intra-molecular cooling in biological systems? -- A quantum thermodynamic perspective.
\newblock \emph{Preprint available at http://arxiv.org/abs/0806.4552}.


\bibitem{Ball04}
Ball, P.
\newblock Enzymes: By chance, or by design?
\newblock \emph{Nature} \textbf{431}, 396-397 (2004).

\bibitem{Tunnelling}
Dutton, P. L. {\em et al.}
\newblock Special Issue on Quantum catalysis in enzymes: beyond the transition state theory paradigm.
\newblock \emph{Phil. Trans. R. Soc. B} \textbf{361}, No.\ 1472, 1293--1455 (2006).


\bibitem{Fleming07}
Engel, G. S., \emph{et al.}
\newblock Evidence for wavelike energy transfer through quantum coherence in photosynthetic systems.
\newblock \emph{Nature} \textbf{446}, 782-786 (2007);

\bibitem{Fleming0702}
Lee, H., \emph{et al.}
\newblock Coherence Dynamics in Photosynthesis: Protein Protection of Excitonic Coherence.
\newblock \emph{Science} \textbf{316}, 1462-1465 (2007).

\bibitem{Mohseni0805}
Mohseni, M. \emph{et al.}
\newblock Environment-assisted quantum walks in photosynthetic energy transfer.
\newblock \emph{J. Chem. Phys.} \textbf{129}, 174106 (2008).


\bibitem{Rebentrost0807}
Rebentrost, P., Mohseni, M., \emph{et al.}
\newblock Environment-Assisted Quantum Transport.
\newblock \emph{Preprint available at http://arxiv.org/abs/0807.0929}.

\bibitem{Plenio0807}
Plenio, M. B. \& Huelga, S. F.
\newblock Dephasing assisted transport: Quantum networks and biomolecules.
\newblock \emph{New J. Phys.} \textbf{10}, 113019 (2008).

\bibitem{Frauenfelder99}
Frauenfelder, H., Wolynes, P. G. \& Austin, R. H.
\newblock Biological Physics.
\newblock \emph{Rev. Mod. Phys.} \textbf{71}, S419-S430 (1999).

\bibitem{Alberts08}
Alberts, B., \emph{et al.}
\newblock \emph{Molecular Biology of the Cell}, (Garland Science, New York, 2008).

\bibitem{Gilmore08}
Gilmore, J. \& McKenzie, R. H.
\newblock Quantum dynamics of electronic excitations in biomolecular chromophores: Role of the protein environment and solvent.
\newblock \emph{J. Phys. Chem. A} \textbf{112}, 2162-2176 (2008).

\bibitem{Hartmann06}
Hartmann, L., D\"ur, W. \& Briegel, H.-J.
\newblock Steady-state entanglement in open and noisy quantum systems
at high temperature.
\newblock \emph{Phys. Rev. A} \textbf{74}, 052304 (2006).

\bibitem{BreuerBook}
Breuer, H.~P. \& Petruccione, F.
\newblock \emph{The Theory of Open Quantum Systems} (Oxford Univerity Press, New York, 2002).

\bibitem{Gilmore07}
Gilmore, J. \& McKenzie, R. H.
\newblock Spin-boson models for quantum decoherence of electronic excitations of biomolecules and quantum dots in a solvent.
\newblock \emph{J. Phys: Cond. Mat.} \textbf{17}, 1735-1746 (2005).

\bibitem{Wootters98}
Wootters, W. K.
\newblock Entanglement of Formation of an Arbitrary State of Two Qubits.
\newblock \emph{Phys. Rev. Lett.} \textbf{80}, 22452-2248 (1998).

\bibitem{Johnson0708}
Olaya-Castro, A., \emph{et al.}
\newblock Efficiency of energy transfer in a light-harvesting system under quantum coherence.
\newblock \emph{Phys. Rev. B} \textbf{78}, 085115 (2008).

\bibitem{Adolphs06}
Adolphs, J. \& Renger, T.
\newblock How Proteins Trigger Excitation Energy Transfer in the FMO Complex of Green Sulfur Bacteria.
\newblock \emph{Biophys. J.} \textbf{91}, 2778-2797 (2006).

\bibitem{Longpaper}
Cai, J.~M., Guerrechi, G.~G., Popescu, S. \& Briegel, H.~J.
\newblock In preparation.

\bibitem{MahlerBook}
Gemmer, J., Michel, M. \& Mahler, G.
\newblock \emph{Quantum Thermodynamics} (Springer Press, New York, 2004).

\bibitem{Fenimore04}
Fenimore, P. W. \emph{et al.}
\newblock Bulk-solvent and hydration-shell fluctuations, similar
to $\alpha$- and $\beta$-fluctuations in glasses, control protein
motions and functions.
\newblock \emph{PNAS} \textbf{101}, 14408–14413 (2004).

\bibitem{Plenio02}
Plenio, M.,\& Huelga, S.
\newblock Entangled light from white noise.
\newblock \emph{Phys. Rev. Lett.} \textbf{88}, 197901 (2002).

\bibitem{Hartmann05}
Hartmann, L. \emph{et al.}
\newblock Spin gases as microscopic models for non-Markovian decoherence.
\newblock \emph{Phys. Rev. A} \textbf{72}, 052107 (2005).



\bibitem{Porras04}
Porras, D. \& Cirac, J. I.
\newblock Effective Quantum Spin Systems with Trapped Ions.
\newblock \emph{Phys. Rev. Lett.} \textbf{92}, 207901 (2004).

\bibitem{Friedenauer08}
Friedenauer, A., \emph{et al.}
\newblock Simulating a quantum magnet with trapped ions.
\newblock \emph{Nature Physics} \textbf{4}, 757-761 (2008).

\bibitem{Wineland0805}
Ospelkaus, C., \emph{et al.}
\newblock Trapped-ion quantum logic gates based on oscillating magnetic fields.
\newblock \emph{Phys. Rev. Lett.} \textbf{101}, 090502 (2008).

\bibitem{Wunderlich}
Wunderlich, Ch.
\newblock Conditional Spin Resonance with Trapped Ions.
\newblock In \emph{Laser Physics at the Limit}, (Springer, New York,
2001).

\bibitem{Poyatos96}
Poyatos, J. F., Cirac, J. I. \& Zoller, P.
\newblock Quantum Reservoir Engineering with Laser Cooled Trapped Ions.
\newblock \emph{Phys. Rev. Lett.} \textbf{77}, 4728-4731 (1996).

\bibitem{RoosPRL}
Roos, C. F., \emph{et al.}
\newblock Bell States of Atoms with Ultralong Lifetimes and Their Tomographic State Analysis.
\newblock \emph{Phys. Rev. Lett.} \textbf{92}, 220402 (2004).

\bibitem{Bouwmeester}
Bouwmeester, D. \newblock{Private discussion}.

\end{thebibliography}

\begin{thebibliography}{9}

\bibitem{BreuerBookSI} Breuer, H.~P. \& Petruccione, F.
\newblock \emph{The Theory of Open Quantum Systems}
(Oxford Univerity Press, New York, 2002).

\bibitem{Hartmann05SI}
Hartmann, L. \emph{et al.}
\newblock Spin gases as microscopic models for non-Markovian decoherence.
\newblock \emph{Phys. Rev. A} \textbf{72}, 052107 (2005).

\bibitem{Hartmann07SI}
Hartmann, L., D\"ur, W. \& Briegel,  H. J.
\newblock Entanglement and its dynamics in open, dissipative systems.
\newblock \emph{New J. Phys.} \textbf{9}, 230 (2007).

\bibitem{Wootters98SI} Wootters, W. K.
\newblock Entanglement of Formation of an Arbitrary State of Two Qubits.
\newblock \emph{Phys. Rev. Lett.} \textbf{80}, 2452-2248 (1998).

\bibitem{LongpaperSI} Cai, J.~M. \emph{et al.}
\newblock In preparation.

\bibitem{Frauenfelder09SI}
Frauenfelder, H. \emph{et al.}
\newblock A unified model of protein dynamics.
\newblock \emph{PNAS} (2009), in press.

\bibitem{Stamp00SI}
Prokov{'}ef, N. \& Stamp, P.
\newblock Theory of the spin bath.
\newblock \emph{Rep. Prog. Phys.} \textbf{63}, 669 (2000).

\bibitem {McK08SI}
Gilmore, J., McKenzie, R. H.
\newblock Quantum dynamics of electronic excitations in biomolecular chromophores: role of the protein environment and
solvent.
\newblock \emph{ J. Phys. Chem. A.} \textbf{112}, 2162 (2008).

\bibitem{Mark95aSI}
Makri, N., Makarov, D. E.
\newblock Tensor propagator for iterative quantum time evolution of reduced density matrices. I.
Theory.
\newblock \emph{J. Chem. Phys.} \textbf{102}, 4600 (1995).

\bibitem{Mark95bSI}
Makri, N., Makarov, D. E.
\newblock Tensor propagator for iterative quantum time evolution of reduced density matrices. II. Numerical
methodology.
\newblock \emph{J. Chem. Phys.} \textbf{102}, 4611 (1995).







\end{thebibliography}
\end{document}